\documentclass[10pt,a4paper,twocolumn,english,prb,aps,showpacs,floatfix,groupedaddress,superscriptaddress,longbibliography]{revtex4-1}
\usepackage{graphicx}
\usepackage{epsfig}
\usepackage[english]{babel}
\usepackage{amsmath}
\usepackage{amssymb}
\usepackage{amsfonts}
\usepackage{longtable}
\usepackage{dcolumn}
\usepackage{bm}
\usepackage{bbm}
\usepackage{nicefrac}
\usepackage{color,array}
\usepackage{colortbl}
\usepackage{dsfont}
\usepackage{mathtools}
\usepackage{subfigure}  
\usepackage{soul}

\usepackage{xcolor,cancel}

\usepackage[breaklinks,colorlinks=true,citecolor=blue,linkcolor=blue]{hyperref}
\usepackage{breakurl}

\begin{document}

\newcommand{\be}{\begin{equation}}
\newcommand{\ee}{\end{equation}}
\newcommand{\bearr}{\begin{eqnarray}}
\newcommand{\eearr}{\end{eqnarray}}
\newcommand{\bseq}{\begin{subequations}}
\newcommand{\eseq}{\end{subequations}}
\newcommand{\nn}{\nonumber}
\newcommand{\dagg}{{\dagger}}
\newcommand{\vpdag}{{\vphantom{\dagger}}}
\newcommand{\bpm}{\begin{pmatrix}} 
\newcommand{\epm}{\end{pmatrix}} 
\newcommand{\bs}{\boldsymbol}

\title{Competing Charge and Magnetic Order in Fermionic Multi-Component Systems}

\author{Mohsen Hafez-Torbati}
\email{torbati@itp.uni-frankfurt.de}
\affiliation{Institut f{\"u}r Theoretische Physik, Goethe-Universit{\"a}t,
60438 Frankfurt/Main, Germany.}

\author{Walter Hofstetter}
\email{hofstett@physik.uni-frankfurt.de}
\affiliation{Institut f{\"u}r Theoretische Physik, Goethe-Universit{\"a}t,
60438 Frankfurt/Main, Germany.}

\date{\today}

\begin{abstract}
We consider the fermionic SU($3$) Hubbard model on the triangular lattice at $1/3$ filling  in the presence 
of a three-sublattice staggered potential which provides the possibility to 
investigate the competition of charge and magnetic order in three-component systems. 
We show that depending on the strength of the staggered potential $\Delta$, the 
Hubbard interaction $U$ destabilizes the band insulator (BI) at small $U$ 
into the Mott insulator (MI) at large $U$ in three different ways with different 
intermediate phases. 
This leads to a rich phase diagram in the $U$-$\Delta$ plane. 
Our results indicate that multi-component systems show not only exotic states 
in the Mott regime as has been considered previously, but also interesting competition between 
charge and magnetic orders which can lead to the emergence of charge-ordered 
magnetic insulators and charge-ordered magnetic metals.
\end{abstract}


\maketitle

\section{introduction}
The observation of Bose-Einstein condensation \cite{Anderson1995} triggered a huge research interest 
in ultracold atoms trapped in optical lattices as flexible and highly controllable quantum simulators 
not only to mimic models of solid state physics but also to study systems which have no obvious
solid state counterparts \cite{Bloch2008,Sugawa2011,Hofstetter2018}. 

Alkali and alkaline-earth-like atoms have up to 
$N=10$ internal states available, which due to the perfect decoupling of the nuclear spin from the 
electronic angular momentum can be used to simulate multi-component systems with 
SU($N$) symmetry \cite{Gorshkov2010a,Cazalilla2014,Ozawa2018}.
Theoretical predictions depending on the value of $N$ suggest multi-component magnetism \cite{Honerkamp2004,Toth2010,Inaba2010,Sotnikov2014,Jakab2016}, 
valence-bond solid states \cite{Jakab2016,Zhou2016,Hermele2011}, and quantum liquids \cite{Hermele2011,Hermele2009,Corboz2012} in the Mott regime. 
A three-component Fermi gas with SU($3$) symmetry has been realized using $^{6}$Li atoms in high magnetic field \cite{Ottenstein2008,Huckans2009} 
and the fermionic SU($6$) Hubbard model has been realized using $^{173}$Yb \cite{Taie2012}. 

In this work we demonstrate that multi-component systems show not only exotic phases in the Mott regime 
as has been discussed previously, but also interesting competition between charge and magnetic order with a possible emergence of 
charge-ordered magnetic metals. 

\section{Model and main results}
\begin{figure}[t]
    \centering
     \includegraphics[width=0.98\linewidth,angle=0]{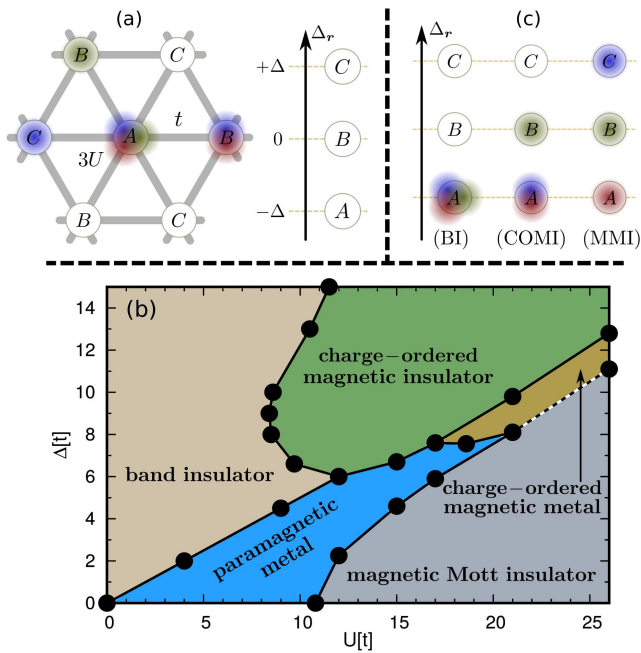}
     \caption{(color online). (a) Schematic representation of the Hamiltonian \eqref{eq:hamiltonian} on 
     the triangular lattice.  The three sublattices $A$, $B$, and $C$ acquire different on-site energies 
     due to the staggered potential $\Delta_{\bs{r}}$. (b) The phase diagram of the model 
     \eqref{eq:hamiltonian} at $1/3$ filling for the inverse temperature $\beta=20/t$ in the $U$-$\Delta$ 
     plane with energies given in units of the hopping parameter $t$, 
     computed using dynamical mean-field theory method. 
     The continuous and the dashed lines denote respectively the second and the first order phase transitions.
     (c) Schematic representation of the 
     different phases: band insulator (BI), 
     where mainly the sublattice $A$ is occupied, charge-ordered magnetic insulator (COMI), where the sublattice 
     $A$ is occupied by two fermionic components and the third component occupies the sublattice $B$, and magnetic 
     Mott insulator (MMI) where each component occupies one of the three sublattices.}
     \label{fig:hamiltonian}
\end{figure}

Our starting point is to introduce a three-sublattice staggered potential into the fermionic SU($3$) Hubbard model 
on the triangular lattice, which allows for the competition of the band insulator (BI) and Mott insulator (MI) phases 
at $1/3$ filling. The Hamiltonian of the system reads
\bearr
\label{eq:hamiltonian}
H\!=&-&t\sum_{\bs r}\sum_{{\bs \delta}} 
\left(
\Psi^\dagg_{\bs{r}+\bs{\delta}} \Psi^\vpdag_{\bs r} + {\rm H.c.}
\right)
+\frac{U}{2}\sum_{\bs r} \! \Psi^\dagg_{\bs{r}} \Psi^\vpdag_{\!\bs{r}}  \Psi^\dagg_{\bs{r}} \Psi^\vpdag_{\!\bs{r}}  \nn \\
&-& \sum_{\bs r} \Delta_{\bs{r}}^\vpdag \Psi^\dagg_{\bs{r}} \Psi^\vpdag_{\bs{r}} \quad,
\eearr
where $\Psi^\dagg_{\bs r}:=\left( c^\dagg_{{\bs r},0}, c^\dagg_{{\bs r}, 1},c^\dagg_{{\bs r}, 2} \right)$ 
is the SU($3$) creation field operator with $c^\dagg_{{\bs r} \alpha}$ being the 
fermionic creation operator at the lattice position ${\bs r}$ with the internal component $\alpha$, 
and $\bs{\delta}$ stands for the nearest-neighbor (NN)
vectors on the triangular lattice. 
The first two terms in Eq. \eqref{eq:hamiltonian} describe the three-component Hubbard model written 
in  SU($3$)-symmetric form, and
the last term is a staggered potential which gives, respectively, the on-site energies $-\Delta$, $0$, and $+\Delta$ 
to the three sublattices $A$, $B$, and $C$ of the triangular lattice, see Fig. \ref{fig:hamiltonian}(a).
Fig. \ref{fig:hamiltonian}(b) displays the phase diagram of the model 
for the inverse temperature $\beta=20/t$ in the $U$-$\Delta$ plane in 
units of the hopping parameter $t$ obtained using the real-space dynamical mean-field theory approach \cite{Potthoff1999}. 
The continuous and the dashed lines correspond respectively to the second and the first order transitions.
Depending on the value of 
$\Delta$, the BI phase
is affected by the Hubbard 
$U$ in different ways. For $0< \Delta \lesssim 6$ the Hubbard interaction drives 
the BI into a paramagnetic metal (PM) and subsequently into a three-sublattice magnetic MI (MMI) with a $120^\circ$ 
pseudospin spiral order \cite{Hafez-Torbati2018}. 
We call the phase ``magnetic'' as it breaks the SU($3$) symmetry, leading 
to a finite expectation value for the pseudospin operator 
$\bs{\mathcal{S}}_{\bs{r}}=\frac{1}{2}\Psi^\dagg_{\bs{r}} {\bs{\lambda}} \Psi^\vpdag_{\!\bs{r}}$ 
where $\bs{\lambda}$ is an eight-dimensional vector made of Gell-Mann matrices. 
Due to the spontaneous breaking of SU($3$) symmetry the state is continuously degenerate. 
The solution lying in the $\hat{\mathcal{S}}_3\!-\!\hat{\mathcal{S}}_8$ plane corresponds to 
a diagonal local density matrix, i.e., $\langle c^\dagg_{{\bs r}\alpha} c^\vpdag_{{\bs r}\beta} \rangle$=0 
for $\alpha\! \neq \! \beta$. In this state at each sublattice one of the 
components has the dominant density \cite{Hafez-Torbati2018}.

For $6t \lesssim \Delta \lesssim 8t$ the Hubbard interaction destabilizes 
the BI into a charge-ordered magnetic insulator (COMI) 
at a first transition point. 
In the COMI phase, sublattices $A$ and $B$ form a $180^\circ$ pseudospin order.
Interestingly, upon further increasing the Hubbard interaction, 
the broken SU($3$) symmetry is restored and the system enters the PM. The transition into the MMI phase 
occurs at a third transition point.
For larger values of the staggered potential, $\Delta\gtrsim8t$, 
the PM is replaced by a charge-ordered magnetic metal (COMM) which 
separates the COMI from the MMI phase. 
We notice that there is a non-uniform charge distribution for any finite value of $\Delta$ 
in the system. The MMI and the PM are not called charge-ordered as they are 
adiabatically connected to the $\Delta=0$ limit where there is a uniform charge distribution.
In contrast, the COMI phase is not equivalent to any phase with a uniform charge distribution 
and the charge-order is a fundamental feature of this state. The same for the COMM phase.

In the limit $U,\Delta\gg t$, the BI-to-COMI transition approaches the line $\Delta \simeq 2U-8t$ and 
the transitions from the COMI to COMM and from COMM to MMI take place, respectively, at $\Delta \simeq U/2$ 
and $\Delta \simeq U/2-2t$. 
This is in perfect agreement with the atomic limit ($t=0$) results.
In the atomic limit one can distinguish the three phases BI, COMI, and MMI depicted 
in Fig. \ref{fig:hamiltonian}(c) with the ground state energies 
$\epsilon_0^{\rm BI}\!=\!U\!-\!\Delta$, $\epsilon_0^{\rm COMI}\!=\!(U\!-\!2\Delta)/3$, and 
$\epsilon_0^{\rm MMI}\!=\!0$ per lattice site.
By comparing these energies one finds that 
BI is stable for $U<\Delta/2$, COMI is stable for $\Delta/2<\!U<\!2\Delta$, 
and MMI is stable for $U\!>\!2\Delta$. This simple atomic limit discussion 
shows how the competition between the staggered potential and 
the Hubbard interaction in fermionic three-component systems can lead to 
the novel COMI phase. The width of the COMM is finite for any finite value of $t$.
We would like to mention that, precisely speaking, the COMI and the MMI phases 
are highly degenerate in the atomic limit and a finite NN hopping is needed to stabilize
the three-sublattice magnetic orders, which can be understood from a second 
order perturbation theory.

\section{Some technical aspects}
The Hamiltonian \eqref{eq:hamiltonian} in the absence of the Hubbard interaction $U$ 
reduces to a three-level problem in momentum space and represents a BI for 
any finite value of $\Delta$. 
In order to investigate the phase diagram of the Hamiltonian \eqref{eq:hamiltonian} 
we employed the dynamical mean-field theory (DMFT) technique  which becomes exact in the limit of
infinite dimensions \cite{Georges1996}.
The method is exact also in the non-interacting and in the atomic limit, and 
by fully taking into account local quantum fluctuations, 
it is a non-perturbative approach for studying the competition of charge and magnetic order in strongly 
correlated systems. 
We use the exact diagonalization impurity solver which enables 
us to compute local quantities with high accuracy, to directly access the real-frequency dynamical spectral functions, 
and to handle the large-$U$ limit with no difficulty. 
The results of ED and hybridization-expansion CTQMC \cite{Gull2011} solver for the finite temperature
phase transitions of the fermionic SU($3$) Hubbard model match nicely \cite{Sotnikov2015}.
We use the real-space DMFT method \cite{Potthoff1999,Snoek2008} 
which we implemented for 
fermionic SU($N$) systems in Ref. \onlinecite{Hafez-Torbati2018}. Due to the absence of electron-hole symmetry we 
add a chemical potential term to the Hamiltonian \eqref{eq:hamiltonian} and adjust it during the DMFT loop to achieve
the desired $1/3$ filling. 
We consider the inverse temperature $\beta=20/t$. 
One notices that the temperature $T=t/20$ is about $10$ 
times smaller than the width of the points chosen in Fig. \ref{fig:hamiltonian}(b) to separate different phases. 
The energy of each state is calculated \cite{Hafez-Torbati2018} and in the coexistence 
regions always the state with the lowest energy is considered as the stable state.

\begin{figure}[t]
    \centering
     \includegraphics[width=1.06\linewidth,angle=-90]{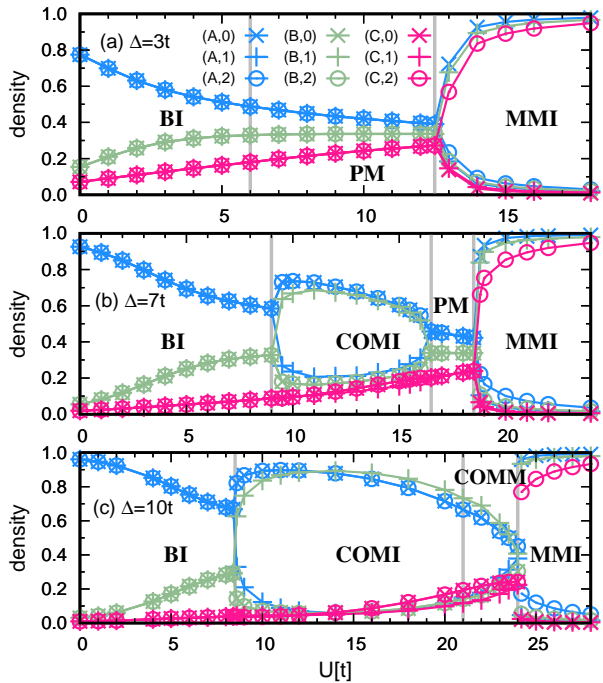}
     \caption{(color online). Local density at the different sublattices $A$, $B$, and $C$ 
     and for the different components $0$, $1$, and $2$ plotted versus the Hubbard interaction $U$ 
     at the staggered potentials $\Delta=3t$ (a), $\Delta=7t$ (b), and $\Delta=10t$ (c). The different 
     phases band insulator (BI), paramagnetic metal (PM), three-sublattice magnetic Mott insulator (MMI), 
     charge-ordered magnetic insulator (COMI), and charge-ordered magnetic metal (COMM) are distinguished. 
     The results shown are obtained for $4$ bath sites of the impurity solver.}
     \label{fig:density}
\end{figure}

\section{density and local moment}
We have plotted the local density 
$\langle c^\dagg_{{\bs r} \alpha} c^\vpdag_{{\bs r} \alpha} \rangle$ on the 
different sublattices $A$, $B$, and $C$ and for the different internal components $\alpha=0,1,2$ 
versus the Hubbard 
$U$ in Fig. \ref{fig:density} for 
$\Delta=3t$ (a), $\Delta=7t$ (b), and $\Delta=10t$ (c). 
The results are obtained for 4 bath sites of the impurity solver.

\begin{figure}[t]
    \centering
     \includegraphics[width=0.75\linewidth,angle=-90]{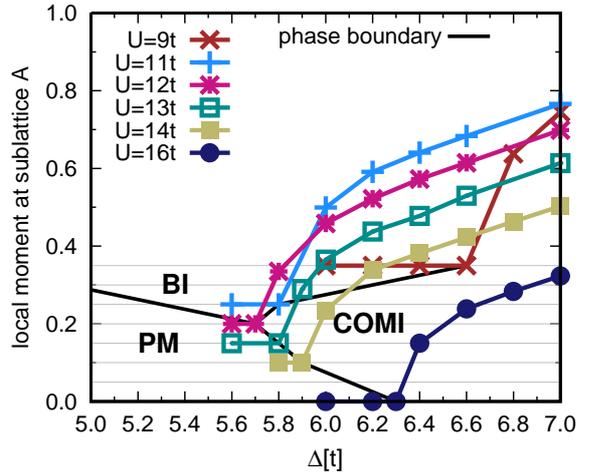}
     \caption{(color online). Local moment on sublattice $A$ plotted versus the staggered potential $\Delta$ 
     for different values of the 
     Hubbard interaction $U$. 
     The local moment is shifted for clarity by $(16t-U)\times0.05$ along the vertical axis.
     We have used BI for band insulator, PM for 
     paramagnetic metal, and COMI for charge-ordered magnetic insulator. The results are for 5 bath sites of the 
     impurity solver.}
     \label{fig:moment}
\end{figure}

One can see from Fig. \ref{fig:density}(a)  that upon increasing the Hubbard interaction $U$ from zero 
in the BI phase the particle density at the sublattice $A$ decreases and the sublattices 
$B$ and $C$ get more populated. The system enters the PM at $U\simeq 6t$, which is signaled 
by a finite density of states at the Fermi energy. 
We notice that due to the finite number of bath sites in the impurity model the fine details of the spectral 
function are not captured and the BI-to-PM transition point is only approximately determined. 
However, we believe that increasing the number of bath sites can not significantly shift the position 
of the predicted transition point.
In the MMI phase for $U \gtrsim 12.5t$, 
each sublattice is mostly occupied with one of the three components.
For the stronger staggered potential $\Delta=7t$ in Fig. \ref{fig:density}(b) 
there is a phase transition at $U\simeq 9t$ from BI into the COMI. 
This phase obviously
shows both magnetic and charge orders. 
In the presence of a weak interaction anisotropy \cite{Sotnikov2014} the component with 
stronger interaction will always occupy the sublattice $B$.
Interestingly, the broken SU($3$) symmetry in the 
COMI phase is  restored again upon increasing the Hubbard interaction to $U\simeq 16.5t$, where the system 
enters the PM. 
It is remarkable that the Hubbard interaction, at least in this particular problem, 
can drive a phase with long-range magnetic order into a PM. 
One notices that the transition from COMI to PM is identified from the local density, for which the 
ED impurity solver is expected to have a high accuracy.
Although the PM-to-MMI transition at $\Delta=7t$ is sharper than the one at $\Delta=3t$ 
it still seems to be continuous. 
In Fig. \ref{fig:hamiltonian}(b) a phase transition is considered second order if the 
local physical quantities such as density and double occupancy change continuously across the 
transition point, and it is considered first order if the change is discontinuous.
Nevertheless, one notices that it is not the aim of the 
present article to discuss the type of phase transitions in the model \eqref{eq:hamiltonian}.
Upon increasing the staggered potential from $\Delta=7t$ to $\Delta=10t$ 
in Fig. \ref{fig:density}(c), the width of the COMI becomes larger, the PM gets substituted with a COMM, 
and the transition to the MMI phase becomes discontinuous. The COMM shows both charge and magnetic orders 
and a finite density of states at the Fermi energy.

One can see from Fig. \ref{fig:density} that for small Hubbard $U$ there is a strong non-uniform 
charge distribution in the system and for large Hubbard $U$ there is a strong magnetic order with an 
almost uniform charge distribution. For intermediate values of $U$ these two different orders compete, leading 
to the emergence of novel phenomena as we discussed above.

The results obtained for 4 and 5 bath sites perfectly agree away from the transition points. However, 
some deviations occur close to the transition points especially near the BI-PM-COMI tricritical point. 
In Fig. \ref{fig:moment} we have plotted the 
local moment $m^\vpdag_{\bs{r}}:=\sqrt{3}|\langle \bs{\mathcal{S}}_{\bs{r}} \rangle|$ at sublattice $A$ obtained for 5 bath sites
versus $\Delta$ for different values of $U$ near the BI-PM-COMI tricritical point. 
The local moment is shifted for clarity by $(16t-U)\times0.05$ along the vertical axis.
In the COMI phase the local moment on sublattice $A$ and on sublattice $B$ 
is the same, while it is zero on sublattice $C$ within our numerical accuracy. We have included the prefactor 
$\sqrt{3}$ in the definition of $m^\vpdag_{\bs{r}}$ in order to have a local moment of $1$ in the fully polarized case, which for the COMI phase occurs when two components occupy 
sublattice $A$, the third component occupies sublattice $B$, and no particle occupies sublattice C. One notices that, 
although there is a small shift in the phase boundaries in Fig. \ref{fig:moment} compared to Fig. \ref{fig:hamiltonian}(b), 
the general shape is the same.

\section{spectral function}
\begin{figure}[t]
    \centering
     \includegraphics[width=1.18\linewidth,angle=-90]{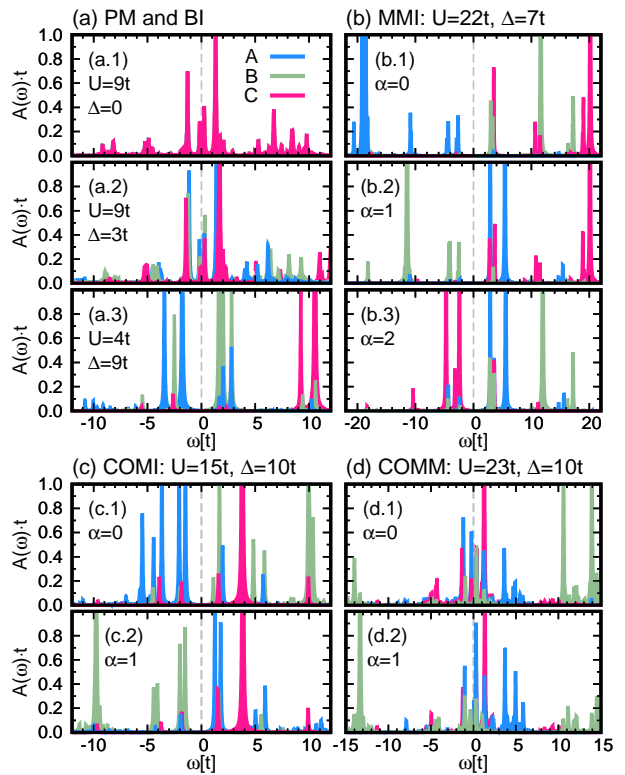}
     \caption{(color online). The spectral function $A(\omega)$ plotted versus energy $\omega$ in 
     the paramagnetic (a) and magnetically ordered phases (b-d). For the paramagnetic metal (PM) 
     and the band insulator (BI) the spectral function is independent than the internal component $\alpha$. 
     For the 3-sublattice magnetic Mott insulator (MMI) the spectral functions of all 
     the three internal components $\alpha=0,1,2$ are represented. 
     For the charge-ordered magnetic insulator (COMI), and the charge-ordered magnetic metal (COMM) 
     the spectral functions of components $\alpha=0$ and $\alpha=2$ are the same due to the symmetry.
     In each panel we have distinguished 
     the spectral functions of the different sublattices $A$, $B$, and $C$ by the different colors 
     blue, green, and red, respectively. The results are for $5$ bath sites in the Anderson impurity 
     problem.}
     \label{fig:spectral}
\end{figure}

Next we discuss the single-particle spectral function, which is given in terms of the 
imaginary part of the single-particle Green's function: 
$A_{\bs{r}\alpha}(\omega)=-\frac{1}{\pi}{\rm Im}G_{\bs{r}\alpha,\bs{r}\alpha}(\omega+i\epsilon)$,
where $\epsilon=0.05$ is the broadening factor.
The spectral function for 5 bath sites in the Anderson impurity problem 
is plotted in Fig. \ref{fig:spectral} for different paramagnetic (a) and magnetically ordered phases (b-d). 
For the paramagnetic phases PM and BI we have plotted the spectral function of only one component.
For the COMI and the COMM the spectral functions of the components 
$\alpha=2$ and $\alpha=0$ are the same due to the symmetry of the phase.
In each panel of Fig. \ref{fig:spectral} we have distinguished the spectral functions 
at the different sublattices $A$, $B$, and $C$ by the different colors blue, green, and red, respectively.

Fig. \ref{fig:spectral}(a.1) depicts the spectral function in the PM for $(U,\Delta)=(9t,0)$. 
Due to the absence of the staggered potential the spectral functions of the different sublattices 
are the same. 
The larger spectral contribution above the Fermi energy $\omega=0$  
is due to the $1/3$ filling.
Keeping the Hubbard interaction $U=9t$ and introducing the staggered potential $\Delta=3t$ in 
Fig. \ref{fig:spectral}(a.2), the system remains still metallic but spectral functions of different 
sublattices become different. For the sublattice $A$ the spectral contributions are transfered from 
above to below the Fermi energy by introducing $\Delta$, while for the sublattice $C$ 
it is the opposite. 
Fig. \ref{fig:spectral}(a.3) shows the spectral function in the BI phase for the parameters 
$(U,\Delta)=(4t,9t)$. The spectral function below the Fermi energy is dominated by the 
contribution from sublattice $A$. 
Right above the Fermi energy, 
there is a noticeable contribution from sublattice $B$. The high energy contributions 
belong mainly to the sublattice $C$. 
Such a spectral structure is expected, as the system is in the BI phase and there should be 
three well-separated bands due to the large staggered potential.

We have plotted the spectral function in the MMI phase for the model parameters $(U,\Delta)=(22t,7t)$ 
in Fig. \ref{fig:spectral}(b). Panels (b.1) to (b.3) correspond to the components 
$\alpha=0$ to $\alpha=2$. 
There is a Mott gap at the Fermi energy and the spectrum below the Fermi energy 
for each component is dominated by the contribution from one of the three sublattices.
This is what one would expect as the system shows a three-sublattice magnetic order. 
The main low-energy peaks in Figs. \ref{fig:spectral}(b.1) to \ref{fig:spectral}(b.3) do not 
occur at the same energies: 
the peak originating from sublattice $A$ appears at much lower energies than the 
one originating from sublattice $C$.
This energy difference is a result of the finite staggered potential in the system, 
which explicitly breaks the translational symmetry of the lattice and gives different on-site 
energies to the different sublattices. In the absence of $\Delta$, the peaks would have the same weight and  
occur at the same energies.

The spectral function in the COMI phase for $(U,\Delta)=(15t,10t)$ is plotted
in Fig. \ref{fig:spectral}(c).
The spectral function of $\alpha=2$ is not shown as it is the same as the 
spectral function of $\alpha=0$.
We observe that the spectral function  below the Fermi energy $\omega=0$ for the component 
$\alpha=0$ is largely governed by the contribution from sublattice $A$. 
The sublattice $B$ contains the major low-energy contributions of the spectral function 
for the component $\alpha=1$. The contributions of the sublattice $C$ to the spectral 
functions mainly lie above the Fermi energy. These results clearly support a phase which has both charge 
and magnetic order and a finite gap at Fermi energy. 
We have displayed the spectral function 
in the COMM for the parameters $(U,\Delta)=(23t,10t)$ in Fig. \ref{fig:spectral}(d). 
There are contributions below 
$\omega=-15t$ mainly from sublattice $A$ 
and contributions  above $\omega=+15t$ mainly from sublattice $C$, which can not be seen in the figure.
Similar to the COMI, the spectral functions of the two components $\alpha=0$ and $\alpha=2$ are the same. 
The main part of the spectral function for all the three components is concentrated near the 
Fermi energy.

\section{summary and outlook}
To summarize, multi-component systems have attracted a lot of attention in recent years 
due to their possible realization in optical lattices and the emergence of exotic states 
in the Mott regime \cite{Gorshkov2010a,Cazalilla2014,Ozawa2018,Taie2012}. We have provided explicit evidence that 
multi-component systems also show interesting competition between charge and magnetic order with the 
possible emergence of charge-ordered magnetic insulators and charge-ordered magnetic metals.
This has not been considered so far, neither experimentally nor theoretically.
This is achieved by introducing a three-sublattice staggered potential to the fermionic SU($3$) Hubbard 
model on the triangular lattice. We show that depending on the strength 
of the staggered potential, different intermediate phases separate the band insulator (BI) at weak 
and the Mott insulator (MI) at strong Hubbard interactions, resulting in a rich phase diagram. 
The fermionic SU($3$) Hubbard model can be realized in optical lattices using  
$^6$Li \cite{Ottenstein2008,Huckans2009} or $^{173}$Yb \cite{Taie2012}, 
and the staggered potential can be created via a triangular superlattice, which 
also produces the Kagome lattice \cite{Jo2012}, or via the digital 
micromirror device, which can be used at single-site level to create different potential landscapes \cite{Liang2010}.
The charge order can be probed by noise correlation measurements \cite{Messer2015} and the 
magnetic order can be detected using a quantum gas microscope \cite{Mazurenko2017}. The excitation spectrum 
can also be measured using spectroscopic techniques such as radio frequency, Raman, and lattice modulation 
spectroscopy \cite{Bloch2008,Messer2015,Joerdens2008,Loida2015}.

We would like to mention that charge and spin order competition in two-component systems 
has been investigated extensively through the ionic Hubbard model (IHM) \cite{Fabrizio1999,Byczuk2009,Ebrahimkhas2012,Jiang2016} 
and the Hubbard model with nearest-neighbor interaction \cite{Nakamura2000,Sandvik2004,Hafez-Torbati2017,Davoudi2006}.  
The 
IHM has recently been realized in optical lattices, 
and charge order \cite{Messer2015} on the honeycomb lattice and different phase transitions 
in one dimension \cite{Loida2017} have been explored. 
Our results motivate similar investigations for higher spin systems, where  
substantially colder Mott insulators are expected at fixed initial entropies 
due to the Pomeranchuk cooling effect \cite{Hazzard2012,Ozawa2018}.
For the two dimensional IHM, there are currently controversial theoretical predictions regarding 
the nature of the intermediate phase(s) separating the BI and MI phases \cite{Hafez-Torbati2016,Paris2007,Kancharla2007}. 
It will be subject to future research to take into account non-local quantum fluctuations and to search 
for new kinds of quantum states in multi-component systems, 
especially near the critical regions in the phase diagram \ref{fig:hamiltonian}(b).

While the phase transitions from paramagnetic metal to magnetic MI and from band insulator to charge-ordered 
magnetic insulator can be described by a local order parameter, there is no local order parameter to describe the 
band insulator to paramagnetic metal and the charge-ordered magnetic insulator to charge-ordered magnetic metal transitions. 
The nature of different types of phase transitions in the model is also a topic which requires further attention in future studies.

It would be also interesting to include spin-orbit coupling into the hopping term in Eq. \eqref{eq:hamiltonian} \cite{Goldman2010} and to study 
SU($3$) topological phases with charge and magnetic order. 
Another important future step is the determination of the finite temperature phase diagram 
and the critical entropies required to reach different magnetically ordered phases of Fig. \ref{fig:hamiltonian}(b) in 
ultracold atoms experiments.

\section*{acknowledgment}
We would like to thank B. Irsigler,  J. Panas, K. Sandholzer, 
C. Weitenberg, and J.-H. Zheng for useful discussions.  
This work was supported by the Deutsche Forschungsgemeinschaft (DFG, German Research Foundation) 
under Project No. 277974659 via Research Unit FOR 2414. 
This work was also supported by the DFG via
the high performance computing center LOEWE-CSC.


\begin{thebibliography}{45}%
\makeatletter
\providecommand \@ifxundefined [1]{%
 \@ifx{#1\undefined}
}%
\providecommand \@ifnum [1]{%
 \ifnum #1\expandafter \@firstoftwo
 \else \expandafter \@secondoftwo
 \fi
}%
\providecommand \@ifx [1]{%
 \ifx #1\expandafter \@firstoftwo
 \else \expandafter \@secondoftwo
 \fi
}%
\providecommand \natexlab [1]{#1}%
\providecommand \enquote  [1]{``#1''}%
\providecommand \bibnamefont  [1]{#1}%
\providecommand \bibfnamefont [1]{#1}%
\providecommand \citenamefont [1]{#1}%
\providecommand \href@noop [0]{\@secondoftwo}%
\providecommand \href [0]{\begingroup \@sanitize@url \@href}%
\providecommand \@href[1]{\@@startlink{#1}\@@href}%
\providecommand \@@href[1]{\endgroup#1\@@endlink}%
\providecommand \@sanitize@url [0]{\catcode `\\12\catcode `\$12\catcode
  `\&12\catcode `\#12\catcode `\^12\catcode `\_12\catcode `\%12\relax}%
\providecommand \@@startlink[1]{}%
\providecommand \@@endlink[0]{}%
\providecommand \url  [0]{\begingroup\@sanitize@url \@url }%
\providecommand \@url [1]{\endgroup\@href {#1}{\urlprefix }}%
\providecommand \urlprefix  [0]{URL }%
\providecommand \Eprint [0]{\href }%
\providecommand \doibase [0]{http://dx.doi.org/}%
\providecommand \selectlanguage [0]{\@gobble}%
\providecommand \bibinfo  [0]{\@secondoftwo}%
\providecommand \bibfield  [0]{\@secondoftwo}%
\providecommand \translation [1]{[#1]}%
\providecommand \BibitemOpen [0]{}%
\providecommand \bibitemStop [0]{}%
\providecommand \bibitemNoStop [0]{.\EOS\space}%
\providecommand \EOS [0]{\spacefactor3000\relax}%
\providecommand \BibitemShut  [1]{\csname bibitem#1\endcsname}%
\let\auto@bib@innerbib\@empty
\bibitem [{\citenamefont {Anderson}\ \emph {et~al.}(1995)\citenamefont
  {Anderson}, \citenamefont {Ensher}, \citenamefont {Matthews}, \citenamefont
  {Wieman},\ and\ \citenamefont {Cornell}}]{Anderson1995}%
  \BibitemOpen
  \bibfield  {author} {\bibinfo {author} {\bibfnamefont {M.~H.}\ \bibnamefont
  {Anderson}}, \bibinfo {author} {\bibfnamefont {J.~R.}\ \bibnamefont
  {Ensher}}, \bibinfo {author} {\bibfnamefont {M.~R.}\ \bibnamefont
  {Matthews}}, \bibinfo {author} {\bibfnamefont {C.~E.}\ \bibnamefont
  {Wieman}}, \ and\ \bibinfo {author} {\bibfnamefont {E.~A.}\ \bibnamefont
  {Cornell}},\ }\href {\doibase 10.1126/science.269.5221.198} {\bibfield
  {journal} {\bibinfo  {journal} {Science}\ }\textbf {\bibinfo {volume}
  {269}},\ \bibinfo {pages} {198} (\bibinfo {year} {1995})}\BibitemShut
  {NoStop}%
\bibitem [{\citenamefont {Bloch}\ \emph {et~al.}(2008)\citenamefont {Bloch},
  \citenamefont {Dalibard},\ and\ \citenamefont {Zwerger}}]{Bloch2008}%
  \BibitemOpen
  \bibfield  {author} {\bibinfo {author} {\bibfnamefont {I.}~\bibnamefont
  {Bloch}}, \bibinfo {author} {\bibfnamefont {J.}~\bibnamefont {Dalibard}}, \
  and\ \bibinfo {author} {\bibfnamefont {W.}~\bibnamefont {Zwerger}},\ }\href
  {\doibase 10.1103/RevModPhys.80.885} {\bibfield  {journal} {\bibinfo
  {journal} {Rev. Mod. Phys.}\ }\textbf {\bibinfo {volume} {80}},\ \bibinfo
  {pages} {885} (\bibinfo {year} {2008})}\BibitemShut {NoStop}%
\bibitem [{\citenamefont {Sugawa}\ \emph {et~al.}(2011)\citenamefont {Sugawa},
  \citenamefont {Inaba}, \citenamefont {Taie}, \citenamefont {Yamazaki},
  \citenamefont {Yamashita},\ and\ \citenamefont {Takahashi}}]{Sugawa2011}%
  \BibitemOpen
  \bibfield  {author} {\bibinfo {author} {\bibfnamefont {S.}~\bibnamefont
  {Sugawa}}, \bibinfo {author} {\bibfnamefont {K.}~\bibnamefont {Inaba}},
  \bibinfo {author} {\bibfnamefont {S.}~\bibnamefont {Taie}}, \bibinfo {author}
  {\bibfnamefont {R.}~\bibnamefont {Yamazaki}}, \bibinfo {author}
  {\bibfnamefont {M.}~\bibnamefont {Yamashita}}, \ and\ \bibinfo {author}
  {\bibfnamefont {Y.}~\bibnamefont {Takahashi}},\ }\href
  {https://doi.org/10.1038/nphys2028} {\bibfield  {journal} {\bibinfo
  {journal} {Nature Physics}\ }\textbf {\bibinfo {volume} {7}},\ \bibinfo
  {pages} {642} (\bibinfo {year} {2011})}\BibitemShut {NoStop}%
\bibitem [{\citenamefont {Hofstetter}\ and\ \citenamefont
  {Qin}(2018)}]{Hofstetter2018}%
  \BibitemOpen
  \bibfield  {author} {\bibinfo {author} {\bibfnamefont {W.}~\bibnamefont
  {Hofstetter}}\ and\ \bibinfo {author} {\bibfnamefont {T.}~\bibnamefont
  {Qin}},\ }\href {http://stacks.iop.org/0953-4075/51/i=8/a=082001} {\bibfield
  {journal} {\bibinfo  {journal} {Journal of Physics B: Atomic, Molecular and
  Optical Physics}\ }\textbf {\bibinfo {volume} {51}},\ \bibinfo {pages}
  {082001} (\bibinfo {year} {2018})}\BibitemShut {NoStop}%
\bibitem [{\citenamefont {Gorshkov}\ \emph {et~al.}(2010)\citenamefont
  {Gorshkov}, \citenamefont {Hermele}, \citenamefont {Gurarie}, \citenamefont
  {Xu}, \citenamefont {Julienne}, \citenamefont {Ye}, \citenamefont {Zoller},
  \citenamefont {Demler}, \citenamefont {Lukin},\ and\ \citenamefont
  {Rey}}]{Gorshkov2010a}%
  \BibitemOpen
  \bibfield  {author} {\bibinfo {author} {\bibfnamefont {A.~V.}\ \bibnamefont
  {Gorshkov}}, \bibinfo {author} {\bibfnamefont {M.}~\bibnamefont {Hermele}},
  \bibinfo {author} {\bibfnamefont {V.}~\bibnamefont {Gurarie}}, \bibinfo
  {author} {\bibfnamefont {C.}~\bibnamefont {Xu}}, \bibinfo {author}
  {\bibfnamefont {P.~S.}\ \bibnamefont {Julienne}}, \bibinfo {author}
  {\bibfnamefont {J.}~\bibnamefont {Ye}}, \bibinfo {author} {\bibfnamefont
  {P.}~\bibnamefont {Zoller}}, \bibinfo {author} {\bibfnamefont
  {E.}~\bibnamefont {Demler}}, \bibinfo {author} {\bibfnamefont {M.~D.}\
  \bibnamefont {Lukin}}, \ and\ \bibinfo {author} {\bibfnamefont {A.~M.}\
  \bibnamefont {Rey}},\ }\href {http://dx.doi.org/10.1038/nphys1535} {\bibfield
   {journal} {\bibinfo  {journal} {Nature Physics}\ }\textbf {\bibinfo {volume}
  {6}},\ \bibinfo {pages} {289} (\bibinfo {year} {2010})}\BibitemShut {NoStop}%
\bibitem [{\citenamefont {Cazalilla}\ and\ \citenamefont
  {Rey}(2014)}]{Cazalilla2014}%
  \BibitemOpen
  \bibfield  {author} {\bibinfo {author} {\bibfnamefont {M.}~\bibnamefont
  {Cazalilla}}\ and\ \bibinfo {author} {\bibfnamefont {A.}~\bibnamefont
  {Rey}},\ }\href {http://stacks.iop.org/0034-4885/77/i=12/a=124401} {\bibfield
   {journal} {\bibinfo  {journal} {Reports on Progress in Physics}\ }\textbf
  {\bibinfo {volume} {77}},\ \bibinfo {pages} {124401} (\bibinfo {year}
  {2014})}\BibitemShut {NoStop}%
\bibitem [{\citenamefont {Ozawa}\ \emph {et~al.}(2018)\citenamefont {Ozawa},
  \citenamefont {Taie}, \citenamefont {Takasu},\ and\ \citenamefont
  {Takahashi}}]{Ozawa2018}%
  \BibitemOpen
  \bibfield  {author} {\bibinfo {author} {\bibfnamefont {H.}~\bibnamefont
  {Ozawa}}, \bibinfo {author} {\bibfnamefont {S.}~\bibnamefont {Taie}},
  \bibinfo {author} {\bibfnamefont {Y.}~\bibnamefont {Takasu}}, \ and\ \bibinfo
  {author} {\bibfnamefont {Y.}~\bibnamefont {Takahashi}},\ }\href {\doibase
  10.1103/PhysRevLett.121.225303} {\bibfield  {journal} {\bibinfo  {journal}
  {Phys. Rev. Lett.}\ }\textbf {\bibinfo {volume} {121}},\ \bibinfo {pages}
  {225303} (\bibinfo {year} {2018})}\BibitemShut {NoStop}%
\bibitem [{\citenamefont {Honerkamp}\ and\ \citenamefont
  {Hofstetter}(2004)}]{Honerkamp2004}%
  \BibitemOpen
  \bibfield  {author} {\bibinfo {author} {\bibfnamefont {C.}~\bibnamefont
  {Honerkamp}}\ and\ \bibinfo {author} {\bibfnamefont {W.}~\bibnamefont
  {Hofstetter}},\ }\href {\doibase 10.1103/PhysRevLett.92.170403} {\bibfield
  {journal} {\bibinfo  {journal} {Phys. Rev. Lett.}\ }\textbf {\bibinfo
  {volume} {92}},\ \bibinfo {pages} {170403} (\bibinfo {year}
  {2004})}\BibitemShut {NoStop}%
\bibitem [{\citenamefont {T\'oth}\ \emph {et~al.}(2010)\citenamefont {T\'oth},
  \citenamefont {L\"auchli}, \citenamefont {Mila},\ and\ \citenamefont
  {Penc}}]{Toth2010}%
  \BibitemOpen
  \bibfield  {author} {\bibinfo {author} {\bibfnamefont {T.~A.}\ \bibnamefont
  {T\'oth}}, \bibinfo {author} {\bibfnamefont {A.~M.}\ \bibnamefont
  {L\"auchli}}, \bibinfo {author} {\bibfnamefont {F.}~\bibnamefont {Mila}}, \
  and\ \bibinfo {author} {\bibfnamefont {K.}~\bibnamefont {Penc}},\ }\href
  {\doibase 10.1103/PhysRevLett.105.265301} {\bibfield  {journal} {\bibinfo
  {journal} {Phys. Rev. Lett.}\ }\textbf {\bibinfo {volume} {105}},\ \bibinfo
  {pages} {265301} (\bibinfo {year} {2010})}\BibitemShut {NoStop}%
\bibitem [{\citenamefont {Inaba}\ \emph {et~al.}(2010)\citenamefont {Inaba},
  \citenamefont {Miyatake},\ and\ \citenamefont {Suga}}]{Inaba2010}%
  \BibitemOpen
  \bibfield  {author} {\bibinfo {author} {\bibfnamefont {K.}~\bibnamefont
  {Inaba}}, \bibinfo {author} {\bibfnamefont {S.-y.}\ \bibnamefont {Miyatake}},
  \ and\ \bibinfo {author} {\bibfnamefont {S.-i.}\ \bibnamefont {Suga}},\
  }\href {\doibase 10.1103/PhysRevA.82.051602} {\bibfield  {journal} {\bibinfo
  {journal} {Phys. Rev. A}\ }\textbf {\bibinfo {volume} {82}},\ \bibinfo
  {pages} {051602} (\bibinfo {year} {2010})}\BibitemShut {NoStop}%
\bibitem [{\citenamefont {Sotnikov}\ and\ \citenamefont
  {Hofstetter}(2014)}]{Sotnikov2014}%
  \BibitemOpen
  \bibfield  {author} {\bibinfo {author} {\bibfnamefont {A.}~\bibnamefont
  {Sotnikov}}\ and\ \bibinfo {author} {\bibfnamefont {W.}~\bibnamefont
  {Hofstetter}},\ }\href {\doibase 10.1103/PhysRevA.89.063601} {\bibfield
  {journal} {\bibinfo  {journal} {Phys. Rev. A}\ }\textbf {\bibinfo {volume}
  {89}},\ \bibinfo {pages} {063601} (\bibinfo {year} {2014})}\BibitemShut
  {NoStop}%
\bibitem [{\citenamefont {Jakab}\ \emph {et~al.}(2016)\citenamefont {Jakab},
  \citenamefont {Szirmai}, \citenamefont {Lewenstein},\ and\ \citenamefont
  {Szirmai}}]{Jakab2016}%
  \BibitemOpen
  \bibfield  {author} {\bibinfo {author} {\bibfnamefont {D.}~\bibnamefont
  {Jakab}}, \bibinfo {author} {\bibfnamefont {E.}~\bibnamefont {Szirmai}},
  \bibinfo {author} {\bibfnamefont {M.}~\bibnamefont {Lewenstein}}, \ and\
  \bibinfo {author} {\bibfnamefont {G.}~\bibnamefont {Szirmai}},\ }\href
  {\doibase 10.1103/PhysRevB.93.064434} {\bibfield  {journal} {\bibinfo
  {journal} {Phys. Rev. B}\ }\textbf {\bibinfo {volume} {93}},\ \bibinfo
  {pages} {064434} (\bibinfo {year} {2016})}\BibitemShut {NoStop}%
\bibitem [{\citenamefont {Zhou}\ \emph {et~al.}(2016)\citenamefont {Zhou},
  \citenamefont {Wang}, \citenamefont {Meng}, \citenamefont {Wang},\ and\
  \citenamefont {Wu}}]{Zhou2016}%
  \BibitemOpen
  \bibfield  {author} {\bibinfo {author} {\bibfnamefont {Z.}~\bibnamefont
  {Zhou}}, \bibinfo {author} {\bibfnamefont {D.}~\bibnamefont {Wang}}, \bibinfo
  {author} {\bibfnamefont {Z.~Y.}\ \bibnamefont {Meng}}, \bibinfo {author}
  {\bibfnamefont {Y.}~\bibnamefont {Wang}}, \ and\ \bibinfo {author}
  {\bibfnamefont {C.}~\bibnamefont {Wu}},\ }\href {\doibase
  10.1103/PhysRevB.93.245157} {\bibfield  {journal} {\bibinfo  {journal} {Phys.
  Rev. B}\ }\textbf {\bibinfo {volume} {93}},\ \bibinfo {pages} {245157}
  (\bibinfo {year} {2016})}\BibitemShut {NoStop}%
\bibitem [{\citenamefont {Hermele}\ and\ \citenamefont
  {Gurarie}(2011)}]{Hermele2011}%
  \BibitemOpen
  \bibfield  {author} {\bibinfo {author} {\bibfnamefont {M.}~\bibnamefont
  {Hermele}}\ and\ \bibinfo {author} {\bibfnamefont {V.}~\bibnamefont
  {Gurarie}},\ }\href {\doibase 10.1103/PhysRevB.84.174441} {\bibfield
  {journal} {\bibinfo  {journal} {Phys. Rev. B}\ }\textbf {\bibinfo {volume}
  {84}},\ \bibinfo {pages} {174441} (\bibinfo {year} {2011})}\BibitemShut
  {NoStop}%
\bibitem [{\citenamefont {Hermele}\ \emph {et~al.}(2009)\citenamefont
  {Hermele}, \citenamefont {Gurarie},\ and\ \citenamefont {Rey}}]{Hermele2009}%
  \BibitemOpen
  \bibfield  {author} {\bibinfo {author} {\bibfnamefont {M.}~\bibnamefont
  {Hermele}}, \bibinfo {author} {\bibfnamefont {V.}~\bibnamefont {Gurarie}}, \
  and\ \bibinfo {author} {\bibfnamefont {A.~M.}\ \bibnamefont {Rey}},\ }\href
  {\doibase 10.1103/PhysRevLett.103.135301} {\bibfield  {journal} {\bibinfo
  {journal} {Phys. Rev. Lett.}\ }\textbf {\bibinfo {volume} {103}},\ \bibinfo
  {pages} {135301} (\bibinfo {year} {2009})}\BibitemShut {NoStop}%
\bibitem [{\citenamefont {Corboz}\ \emph {et~al.}(2012)\citenamefont {Corboz},
  \citenamefont {Lajk\'o}, \citenamefont {L\"auchli}, \citenamefont {Penc},\
  and\ \citenamefont {Mila}}]{Corboz2012}%
  \BibitemOpen
  \bibfield  {author} {\bibinfo {author} {\bibfnamefont {P.}~\bibnamefont
  {Corboz}}, \bibinfo {author} {\bibfnamefont {M.}~\bibnamefont {Lajk\'o}},
  \bibinfo {author} {\bibfnamefont {A.~M.}\ \bibnamefont {L\"auchli}}, \bibinfo
  {author} {\bibfnamefont {K.}~\bibnamefont {Penc}}, \ and\ \bibinfo {author}
  {\bibfnamefont {F.}~\bibnamefont {Mila}},\ }\href {\doibase
  10.1103/PhysRevX.2.041013} {\bibfield  {journal} {\bibinfo  {journal} {Phys.
  Rev. X}\ }\textbf {\bibinfo {volume} {2}},\ \bibinfo {pages} {041013}
  (\bibinfo {year} {2012})}\BibitemShut {NoStop}%
\bibitem [{\citenamefont {Ottenstein}\ \emph {et~al.}(2008)\citenamefont
  {Ottenstein}, \citenamefont {Lompe}, \citenamefont {Kohnen}, \citenamefont
  {Wenz},\ and\ \citenamefont {Jochim}}]{Ottenstein2008}%
  \BibitemOpen
  \bibfield  {author} {\bibinfo {author} {\bibfnamefont {T.~B.}\ \bibnamefont
  {Ottenstein}}, \bibinfo {author} {\bibfnamefont {T.}~\bibnamefont {Lompe}},
  \bibinfo {author} {\bibfnamefont {M.}~\bibnamefont {Kohnen}}, \bibinfo
  {author} {\bibfnamefont {A.~N.}\ \bibnamefont {Wenz}}, \ and\ \bibinfo
  {author} {\bibfnamefont {S.}~\bibnamefont {Jochim}},\ }\href {\doibase
  10.1103/PhysRevLett.101.203202} {\bibfield  {journal} {\bibinfo  {journal}
  {Phys. Rev. Lett.}\ }\textbf {\bibinfo {volume} {101}},\ \bibinfo {pages}
  {203202} (\bibinfo {year} {2008})}\BibitemShut {NoStop}%
\bibitem [{\citenamefont {Huckans}\ \emph {et~al.}(2009)\citenamefont
  {Huckans}, \citenamefont {Williams}, \citenamefont {Hazlett}, \citenamefont
  {Stites},\ and\ \citenamefont {O'Hara}}]{Huckans2009}%
  \BibitemOpen
  \bibfield  {author} {\bibinfo {author} {\bibfnamefont {J.~H.}\ \bibnamefont
  {Huckans}}, \bibinfo {author} {\bibfnamefont {J.~R.}\ \bibnamefont
  {Williams}}, \bibinfo {author} {\bibfnamefont {E.~L.}\ \bibnamefont
  {Hazlett}}, \bibinfo {author} {\bibfnamefont {R.~W.}\ \bibnamefont {Stites}},
  \ and\ \bibinfo {author} {\bibfnamefont {K.~M.}\ \bibnamefont {O'Hara}},\
  }\href {\doibase 10.1103/PhysRevLett.102.165302} {\bibfield  {journal}
  {\bibinfo  {journal} {Phys. Rev. Lett.}\ }\textbf {\bibinfo {volume} {102}},\
  \bibinfo {pages} {165302} (\bibinfo {year} {2009})}\BibitemShut {NoStop}%
\bibitem [{\citenamefont {Taie}\ \emph {et~al.}(2012)\citenamefont {Taie},
  \citenamefont {Yamazaki}, \citenamefont {Sugawa},\ and\ \citenamefont
  {Takahashi}}]{Taie2012}%
  \BibitemOpen
  \bibfield  {author} {\bibinfo {author} {\bibfnamefont {S.}~\bibnamefont
  {Taie}}, \bibinfo {author} {\bibfnamefont {R.}~\bibnamefont {Yamazaki}},
  \bibinfo {author} {\bibfnamefont {S.}~\bibnamefont {Sugawa}}, \ and\ \bibinfo
  {author} {\bibfnamefont {Y.}~\bibnamefont {Takahashi}},\ }\href
  {http://dx.doi.org/10.1038/nphys2430} {\bibfield  {journal} {\bibinfo
  {journal} {Nature Physics}\ }\textbf {\bibinfo {volume} {8}},\ \bibinfo
  {pages} {825} (\bibinfo {year} {2012})}\BibitemShut {NoStop}%
\bibitem [{\citenamefont {Potthoff}\ and\ \citenamefont
  {Nolting}(1999)}]{Potthoff1999}%
  \BibitemOpen
  \bibfield  {author} {\bibinfo {author} {\bibfnamefont {M.}~\bibnamefont
  {Potthoff}}\ and\ \bibinfo {author} {\bibfnamefont {W.}~\bibnamefont
  {Nolting}},\ }\href {\doibase 10.1103/PhysRevB.59.2549} {\bibfield  {journal}
  {\bibinfo  {journal} {Phys. Rev. B}\ }\textbf {\bibinfo {volume} {59}},\
  \bibinfo {pages} {2549} (\bibinfo {year} {1999})}\BibitemShut {NoStop}%
\bibitem [{\citenamefont {Hafez-Torbati}\ and\ \citenamefont
  {Hofstetter}(2018)}]{Hafez-Torbati2018}%
  \BibitemOpen
  \bibfield  {author} {\bibinfo {author} {\bibfnamefont {M.}~\bibnamefont
  {Hafez-Torbati}}\ and\ \bibinfo {author} {\bibfnamefont {W.}~\bibnamefont
  {Hofstetter}},\ }\href {\doibase 10.1103/PhysRevB.98.245131} {\bibfield
  {journal} {\bibinfo  {journal} {Phys. Rev. B}\ }\textbf {\bibinfo {volume}
  {98}},\ \bibinfo {pages} {245131} (\bibinfo {year} {2018})}\BibitemShut
  {NoStop}%
\bibitem [{\citenamefont {Georges}\ \emph {et~al.}(1996)\citenamefont
  {Georges}, \citenamefont {Kotliar}, \citenamefont {Krauth},\ and\
  \citenamefont {Rozenberg}}]{Georges1996}%
  \BibitemOpen
  \bibfield  {author} {\bibinfo {author} {\bibfnamefont {A.}~\bibnamefont
  {Georges}}, \bibinfo {author} {\bibfnamefont {G.}~\bibnamefont {Kotliar}},
  \bibinfo {author} {\bibfnamefont {W.}~\bibnamefont {Krauth}}, \ and\ \bibinfo
  {author} {\bibfnamefont {M.~J.}\ \bibnamefont {Rozenberg}},\ }\href {\doibase
  10.1103/RevModPhys.68.13} {\bibfield  {journal} {\bibinfo  {journal} {Rev.
  Mod. Phys.}\ }\textbf {\bibinfo {volume} {68}},\ \bibinfo {pages} {13}
  (\bibinfo {year} {1996})}\BibitemShut {NoStop}%
\bibitem [{\citenamefont {Gull}\ \emph {et~al.}(2011)\citenamefont {Gull},
  \citenamefont {Millis}, \citenamefont {Lichtenstein}, \citenamefont
  {Rubtsov}, \citenamefont {Troyer},\ and\ \citenamefont {Werner}}]{Gull2011}%
  \BibitemOpen
  \bibfield  {author} {\bibinfo {author} {\bibfnamefont {E.}~\bibnamefont
  {Gull}}, \bibinfo {author} {\bibfnamefont {A.~J.}\ \bibnamefont {Millis}},
  \bibinfo {author} {\bibfnamefont {A.~I.}\ \bibnamefont {Lichtenstein}},
  \bibinfo {author} {\bibfnamefont {A.~N.}\ \bibnamefont {Rubtsov}}, \bibinfo
  {author} {\bibfnamefont {M.}~\bibnamefont {Troyer}}, \ and\ \bibinfo {author}
  {\bibfnamefont {P.}~\bibnamefont {Werner}},\ }\href {\doibase
  10.1103/RevModPhys.83.349} {\bibfield  {journal} {\bibinfo  {journal} {Rev.
  Mod. Phys.}\ }\textbf {\bibinfo {volume} {83}},\ \bibinfo {pages} {349}
  (\bibinfo {year} {2011})}\BibitemShut {NoStop}%
\bibitem [{\citenamefont {Sotnikov}(2015)}]{Sotnikov2015}%
  \BibitemOpen
  \bibfield  {author} {\bibinfo {author} {\bibfnamefont {A.}~\bibnamefont
  {Sotnikov}},\ }\href {\doibase 10.1103/PhysRevA.92.023633} {\bibfield
  {journal} {\bibinfo  {journal} {Phys. Rev. A}\ }\textbf {\bibinfo {volume}
  {92}},\ \bibinfo {pages} {023633} (\bibinfo {year} {2015})}\BibitemShut
  {NoStop}%
\bibitem [{\citenamefont {Snoek}\ \emph {et~al.}(2008)\citenamefont {Snoek},
  \citenamefont {Titvinidze}, \citenamefont {Tőke}, \citenamefont {Byczuk},\
  and\ \citenamefont {Hofstetter}}]{Snoek2008}%
  \BibitemOpen
  \bibfield  {author} {\bibinfo {author} {\bibfnamefont {M.}~\bibnamefont
  {Snoek}}, \bibinfo {author} {\bibfnamefont {I.}~\bibnamefont {Titvinidze}},
  \bibinfo {author} {\bibfnamefont {C.}~\bibnamefont {Tőke}}, \bibinfo
  {author} {\bibfnamefont {K.}~\bibnamefont {Byczuk}}, \ and\ \bibinfo {author}
  {\bibfnamefont {W.}~\bibnamefont {Hofstetter}},\ }\href
  {http://stacks.iop.org/1367-2630/10/i=9/a=093008} {\bibfield  {journal}
  {\bibinfo  {journal} {New Journal of Physics}\ }\textbf {\bibinfo {volume}
  {10}},\ \bibinfo {pages} {093008} (\bibinfo {year} {2008})}\BibitemShut
  {NoStop}%
\bibitem [{\citenamefont {Jo}\ \emph {et~al.}(2012)\citenamefont {Jo},
  \citenamefont {Guzman}, \citenamefont {Thomas}, \citenamefont {Hosur},
  \citenamefont {Vishwanath},\ and\ \citenamefont {Stamper-Kurn}}]{Jo2012}%
  \BibitemOpen
  \bibfield  {author} {\bibinfo {author} {\bibfnamefont {G.-B.}\ \bibnamefont
  {Jo}}, \bibinfo {author} {\bibfnamefont {J.}~\bibnamefont {Guzman}}, \bibinfo
  {author} {\bibfnamefont {C.~K.}\ \bibnamefont {Thomas}}, \bibinfo {author}
  {\bibfnamefont {P.}~\bibnamefont {Hosur}}, \bibinfo {author} {\bibfnamefont
  {A.}~\bibnamefont {Vishwanath}}, \ and\ \bibinfo {author} {\bibfnamefont
  {D.~M.}\ \bibnamefont {Stamper-Kurn}},\ }\href {\doibase
  10.1103/PhysRevLett.108.045305} {\bibfield  {journal} {\bibinfo  {journal}
  {Phys. Rev. Lett.}\ }\textbf {\bibinfo {volume} {108}},\ \bibinfo {pages}
  {045305} (\bibinfo {year} {2012})}\BibitemShut {NoStop}%
\bibitem [{\citenamefont {Liang}\ \emph {et~al.}(2010)\citenamefont {Liang},
  \citenamefont {Rudolph N.~Kohn}, \citenamefont {Becker},\ and\ \citenamefont
  {Heinzen}}]{Liang2010}%
  \BibitemOpen
  \bibfield  {author} {\bibinfo {author} {\bibfnamefont {J.}~\bibnamefont
  {Liang}}, \bibinfo {author} {\bibfnamefont {J.}~\bibnamefont {Rudolph
  N.~Kohn}}, \bibinfo {author} {\bibfnamefont {M.~F.}\ \bibnamefont {Becker}},
  \ and\ \bibinfo {author} {\bibfnamefont {D.~J.}\ \bibnamefont {Heinzen}},\
  }\href {\doibase 10.1364/AO.49.001323} {\bibfield  {journal} {\bibinfo
  {journal} {Appl. Opt.}\ }\textbf {\bibinfo {volume} {49}},\ \bibinfo {pages}
  {1323} (\bibinfo {year} {2010})}\BibitemShut {NoStop}%
\bibitem [{\citenamefont {Messer}\ \emph {et~al.}(2015)\citenamefont {Messer},
  \citenamefont {Desbuquois}, \citenamefont {Uehlinger}, \citenamefont {Jotzu},
  \citenamefont {Huber}, \citenamefont {Greif},\ and\ \citenamefont
  {Esslinger}}]{Messer2015}%
  \BibitemOpen
  \bibfield  {author} {\bibinfo {author} {\bibfnamefont {M.}~\bibnamefont
  {Messer}}, \bibinfo {author} {\bibfnamefont {R.}~\bibnamefont {Desbuquois}},
  \bibinfo {author} {\bibfnamefont {T.}~\bibnamefont {Uehlinger}}, \bibinfo
  {author} {\bibfnamefont {G.}~\bibnamefont {Jotzu}}, \bibinfo {author}
  {\bibfnamefont {S.}~\bibnamefont {Huber}}, \bibinfo {author} {\bibfnamefont
  {D.}~\bibnamefont {Greif}}, \ and\ \bibinfo {author} {\bibfnamefont
  {T.}~\bibnamefont {Esslinger}},\ }\href {\doibase
  10.1103/PhysRevLett.115.115303} {\bibfield  {journal} {\bibinfo  {journal}
  {Phys. Rev. Lett.}\ }\textbf {\bibinfo {volume} {115}},\ \bibinfo {pages}
  {115303} (\bibinfo {year} {2015})}\BibitemShut {NoStop}%
\bibitem [{\citenamefont {Mazurenko}\ \emph {et~al.}(2017)\citenamefont
  {Mazurenko}, \citenamefont {Chiu}, \citenamefont {Ji}, \citenamefont
  {Parsons}, \citenamefont {Kanász-Nagy}, \citenamefont {Schmidt},
  \citenamefont {Grusdt}, \citenamefont {Demler}, \citenamefont {Greif},\ and\
  \citenamefont {Greiner}}]{Mazurenko2017}%
  \BibitemOpen
  \bibfield  {author} {\bibinfo {author} {\bibfnamefont {A.}~\bibnamefont
  {Mazurenko}}, \bibinfo {author} {\bibfnamefont {C.~S.}\ \bibnamefont {Chiu}},
  \bibinfo {author} {\bibfnamefont {G.}~\bibnamefont {Ji}}, \bibinfo {author}
  {\bibfnamefont {M.~F.}\ \bibnamefont {Parsons}}, \bibinfo {author}
  {\bibfnamefont {M.}~\bibnamefont {Kanász-Nagy}}, \bibinfo {author}
  {\bibfnamefont {R.}~\bibnamefont {Schmidt}}, \bibinfo {author} {\bibfnamefont
  {F.}~\bibnamefont {Grusdt}}, \bibinfo {author} {\bibfnamefont
  {E.}~\bibnamefont {Demler}}, \bibinfo {author} {\bibfnamefont
  {D.}~\bibnamefont {Greif}}, \ and\ \bibinfo {author} {\bibfnamefont
  {M.}~\bibnamefont {Greiner}},\ }\href {https://doi.org/10.1038/nature22362}
  {\bibfield  {journal} {\bibinfo  {journal} {Nature}\ }\textbf {\bibinfo
  {volume} {545}},\ \bibinfo {pages} {462} (\bibinfo {year}
  {2017})}\BibitemShut {NoStop}%
\bibitem [{\citenamefont {Jördens}\ \emph {et~al.}(2008)\citenamefont
  {Jördens}, \citenamefont {Strohmaier}, \citenamefont {Günter},
  \citenamefont {Moritz},\ and\ \citenamefont {Esslinger}}]{Joerdens2008}%
  \BibitemOpen
  \bibfield  {author} {\bibinfo {author} {\bibfnamefont {R.}~\bibnamefont
  {Jördens}}, \bibinfo {author} {\bibfnamefont {N.}~\bibnamefont
  {Strohmaier}}, \bibinfo {author} {\bibfnamefont {K.}~\bibnamefont {Günter}},
  \bibinfo {author} {\bibfnamefont {H.}~\bibnamefont {Moritz}}, \ and\ \bibinfo
  {author} {\bibfnamefont {T.}~\bibnamefont {Esslinger}},\ }\href
  {https://doi.org/10.1038/nature07244} {\bibfield  {journal} {\bibinfo
  {journal} {Nature}\ }\textbf {\bibinfo {volume} {455}},\ \bibinfo {pages}
  {204} (\bibinfo {year} {2008})}\BibitemShut {NoStop}%
\bibitem [{\citenamefont {Loida}\ \emph {et~al.}(2015)\citenamefont {Loida},
  \citenamefont {Sheikhan},\ and\ \citenamefont {Kollath}}]{Loida2015}%
  \BibitemOpen
  \bibfield  {author} {\bibinfo {author} {\bibfnamefont {K.}~\bibnamefont
  {Loida}}, \bibinfo {author} {\bibfnamefont {A.}~\bibnamefont {Sheikhan}}, \
  and\ \bibinfo {author} {\bibfnamefont {C.}~\bibnamefont {Kollath}},\ }\href
  {\doibase 10.1103/PhysRevA.92.043624} {\bibfield  {journal} {\bibinfo
  {journal} {Phys. Rev. A}\ }\textbf {\bibinfo {volume} {92}},\ \bibinfo
  {pages} {043624} (\bibinfo {year} {2015})}\BibitemShut {NoStop}%
\bibitem [{\citenamefont {Fabrizio}\ \emph {et~al.}(1999)\citenamefont
  {Fabrizio}, \citenamefont {Gogolin},\ and\ \citenamefont
  {Nersesyan}}]{Fabrizio1999}%
  \BibitemOpen
  \bibfield  {author} {\bibinfo {author} {\bibfnamefont {M.}~\bibnamefont
  {Fabrizio}}, \bibinfo {author} {\bibfnamefont {A.~O.}\ \bibnamefont
  {Gogolin}}, \ and\ \bibinfo {author} {\bibfnamefont {A.~A.}\ \bibnamefont
  {Nersesyan}},\ }\href {\doibase 10.1103/PhysRevLett.83.2014} {\bibfield
  {journal} {\bibinfo  {journal} {Phys. Rev. Lett.}\ }\textbf {\bibinfo
  {volume} {83}},\ \bibinfo {pages} {2014} (\bibinfo {year}
  {1999})}\BibitemShut {NoStop}%
\bibitem [{\citenamefont {Byczuk}\ \emph {et~al.}(2009)\citenamefont {Byczuk},
  \citenamefont {Sekania}, \citenamefont {Hofstetter},\ and\ \citenamefont
  {Kampf}}]{Byczuk2009}%
  \BibitemOpen
  \bibfield  {author} {\bibinfo {author} {\bibfnamefont {K.}~\bibnamefont
  {Byczuk}}, \bibinfo {author} {\bibfnamefont {M.}~\bibnamefont {Sekania}},
  \bibinfo {author} {\bibfnamefont {W.}~\bibnamefont {Hofstetter}}, \ and\
  \bibinfo {author} {\bibfnamefont {A.~P.}\ \bibnamefont {Kampf}},\ }\href
  {\doibase 10.1103/PhysRevB.79.121103} {\bibfield  {journal} {\bibinfo
  {journal} {Phys. Rev. B}\ }\textbf {\bibinfo {volume} {79}},\ \bibinfo
  {pages} {121103} (\bibinfo {year} {2009})}\BibitemShut {NoStop}%
\bibitem [{\citenamefont {Ebrahimkhas}\ and\ \citenamefont
  {Jafari}(2012)}]{Ebrahimkhas2012}%
  \BibitemOpen
  \bibfield  {author} {\bibinfo {author} {\bibfnamefont {M.}~\bibnamefont
  {Ebrahimkhas}}\ and\ \bibinfo {author} {\bibfnamefont {S.~A.}\ \bibnamefont
  {Jafari}},\ }\href {http://stacks.iop.org/0295-5075/98/i=2/a=27009}
  {\bibfield  {journal} {\bibinfo  {journal} {EPL (Europhysics Letters)}\
  }\textbf {\bibinfo {volume} {98}},\ \bibinfo {pages} {27009} (\bibinfo {year}
  {2012})}\BibitemShut {NoStop}%
\bibitem [{\citenamefont {Jiang}\ and\ \citenamefont
  {Schulthess}(2016)}]{Jiang2016}%
  \BibitemOpen
  \bibfield  {author} {\bibinfo {author} {\bibfnamefont {M.}~\bibnamefont
  {Jiang}}\ and\ \bibinfo {author} {\bibfnamefont {T.~C.}\ \bibnamefont
  {Schulthess}},\ }\href {\doibase 10.1103/PhysRevB.93.165146} {\bibfield
  {journal} {\bibinfo  {journal} {Phys. Rev. B}\ }\textbf {\bibinfo {volume}
  {93}},\ \bibinfo {pages} {165146} (\bibinfo {year} {2016})}\BibitemShut
  {NoStop}%
\bibitem [{\citenamefont {Nakamura}(2000)}]{Nakamura2000}%
  \BibitemOpen
  \bibfield  {author} {\bibinfo {author} {\bibfnamefont {M.}~\bibnamefont
  {Nakamura}},\ }\href {\doibase 10.1103/PhysRevB.61.16377} {\bibfield
  {journal} {\bibinfo  {journal} {Phys. Rev. B}\ }\textbf {\bibinfo {volume}
  {61}},\ \bibinfo {pages} {16377} (\bibinfo {year} {2000})}\BibitemShut
  {NoStop}%
\bibitem [{\citenamefont {Sandvik}\ \emph {et~al.}(2004)\citenamefont
  {Sandvik}, \citenamefont {Balents},\ and\ \citenamefont
  {Campbell}}]{Sandvik2004}%
  \BibitemOpen
  \bibfield  {author} {\bibinfo {author} {\bibfnamefont {A.~W.}\ \bibnamefont
  {Sandvik}}, \bibinfo {author} {\bibfnamefont {L.}~\bibnamefont {Balents}}, \
  and\ \bibinfo {author} {\bibfnamefont {D.~K.}\ \bibnamefont {Campbell}},\
  }\href {\doibase 10.1103/PhysRevLett.92.236401} {\bibfield  {journal}
  {\bibinfo  {journal} {Phys. Rev. Lett.}\ }\textbf {\bibinfo {volume} {92}},\
  \bibinfo {pages} {236401} (\bibinfo {year} {2004})}\BibitemShut {NoStop}%
\bibitem [{\citenamefont {Hafez-Torbati}\ and\ \citenamefont
  {Uhrig}(2017)}]{Hafez-Torbati2017}%
  \BibitemOpen
  \bibfield  {author} {\bibinfo {author} {\bibfnamefont {M.}~\bibnamefont
  {Hafez-Torbati}}\ and\ \bibinfo {author} {\bibfnamefont {G.~S.}\ \bibnamefont
  {Uhrig}},\ }\href {\doibase 10.1103/PhysRevB.96.125129} {\bibfield  {journal}
  {\bibinfo  {journal} {Phys. Rev. B}\ }\textbf {\bibinfo {volume} {96}},\
  \bibinfo {pages} {125129} (\bibinfo {year} {2017})}\BibitemShut {NoStop}%
\bibitem [{\citenamefont {Davoudi}\ and\ \citenamefont
  {Tremblay}(2006)}]{Davoudi2006}%
  \BibitemOpen
  \bibfield  {author} {\bibinfo {author} {\bibfnamefont {B.}~\bibnamefont
  {Davoudi}}\ and\ \bibinfo {author} {\bibfnamefont {A.-M.~S.}\ \bibnamefont
  {Tremblay}},\ }\href {\doibase 10.1103/PhysRevB.74.035113} {\bibfield
  {journal} {\bibinfo  {journal} {Phys. Rev. B}\ }\textbf {\bibinfo {volume}
  {74}},\ \bibinfo {pages} {035113} (\bibinfo {year} {2006})}\BibitemShut
  {NoStop}%
\bibitem [{\citenamefont {Loida}\ \emph {et~al.}(2017)\citenamefont {Loida},
  \citenamefont {Bernier}, \citenamefont {Citro}, \citenamefont {Orignac},\
  and\ \citenamefont {Kollath}}]{Loida2017}%
  \BibitemOpen
  \bibfield  {author} {\bibinfo {author} {\bibfnamefont {K.}~\bibnamefont
  {Loida}}, \bibinfo {author} {\bibfnamefont {J.-S.}\ \bibnamefont {Bernier}},
  \bibinfo {author} {\bibfnamefont {R.}~\bibnamefont {Citro}}, \bibinfo
  {author} {\bibfnamefont {E.}~\bibnamefont {Orignac}}, \ and\ \bibinfo
  {author} {\bibfnamefont {C.}~\bibnamefont {Kollath}},\ }\href {\doibase
  10.1103/PhysRevLett.119.230403} {\bibfield  {journal} {\bibinfo  {journal}
  {Phys. Rev. Lett.}\ }\textbf {\bibinfo {volume} {119}},\ \bibinfo {pages}
  {230403} (\bibinfo {year} {2017})}\BibitemShut {NoStop}%
\bibitem [{\citenamefont {Hazzard}\ \emph {et~al.}(2012)\citenamefont
  {Hazzard}, \citenamefont {Gurarie}, \citenamefont {Hermele},\ and\
  \citenamefont {Rey}}]{Hazzard2012}%
  \BibitemOpen
  \bibfield  {author} {\bibinfo {author} {\bibfnamefont {K.~R.~A.}\
  \bibnamefont {Hazzard}}, \bibinfo {author} {\bibfnamefont {V.}~\bibnamefont
  {Gurarie}}, \bibinfo {author} {\bibfnamefont {M.}~\bibnamefont {Hermele}}, \
  and\ \bibinfo {author} {\bibfnamefont {A.~M.}\ \bibnamefont {Rey}},\ }\href
  {\doibase 10.1103/PhysRevA.85.041604} {\bibfield  {journal} {\bibinfo
  {journal} {Phys. Rev. A}\ }\textbf {\bibinfo {volume} {85}},\ \bibinfo
  {pages} {041604} (\bibinfo {year} {2012})}\BibitemShut {NoStop}%
\bibitem [{\citenamefont {Hafez-Torbati}\ and\ \citenamefont
  {Uhrig}(2016)}]{Hafez-Torbati2016}%
  \BibitemOpen
  \bibfield  {author} {\bibinfo {author} {\bibfnamefont {M.}~\bibnamefont
  {Hafez-Torbati}}\ and\ \bibinfo {author} {\bibfnamefont {G.~S.}\ \bibnamefont
  {Uhrig}},\ }\href {\doibase 10.1103/PhysRevB.93.195128} {\bibfield  {journal}
  {\bibinfo  {journal} {Phys. Rev. B}\ }\textbf {\bibinfo {volume} {93}},\
  \bibinfo {pages} {195128} (\bibinfo {year} {2016})}\BibitemShut {NoStop}%
\bibitem [{\citenamefont {Paris}\ \emph {et~al.}(2007)\citenamefont {Paris},
  \citenamefont {Bouadim}, \citenamefont {Hebert}, \citenamefont {Batrouni},\
  and\ \citenamefont {Scalettar}}]{Paris2007}%
  \BibitemOpen
  \bibfield  {author} {\bibinfo {author} {\bibfnamefont {N.}~\bibnamefont
  {Paris}}, \bibinfo {author} {\bibfnamefont {K.}~\bibnamefont {Bouadim}},
  \bibinfo {author} {\bibfnamefont {F.}~\bibnamefont {Hebert}}, \bibinfo
  {author} {\bibfnamefont {G.~G.}\ \bibnamefont {Batrouni}}, \ and\ \bibinfo
  {author} {\bibfnamefont {R.~T.}\ \bibnamefont {Scalettar}},\ }\href {\doibase
  10.1103/PhysRevLett.98.046403} {\bibfield  {journal} {\bibinfo  {journal}
  {Phys. Rev. Lett.}\ }\textbf {\bibinfo {volume} {98}},\ \bibinfo {pages}
  {046403} (\bibinfo {year} {2007})}\BibitemShut {NoStop}%
\bibitem [{\citenamefont {Kancharla}\ and\ \citenamefont
  {Dagotto}(2007)}]{Kancharla2007}%
  \BibitemOpen
  \bibfield  {author} {\bibinfo {author} {\bibfnamefont {S.~S.}\ \bibnamefont
  {Kancharla}}\ and\ \bibinfo {author} {\bibfnamefont {E.}~\bibnamefont
  {Dagotto}},\ }\href {\doibase 10.1103/PhysRevLett.98.016402} {\bibfield
  {journal} {\bibinfo  {journal} {Phys. Rev. Lett.}\ }\textbf {\bibinfo
  {volume} {98}},\ \bibinfo {pages} {016402} (\bibinfo {year}
  {2007})}\BibitemShut {NoStop}%
\bibitem [{\citenamefont {Goldman}\ \emph {et~al.}(2010)\citenamefont
  {Goldman}, \citenamefont {Satija}, \citenamefont {Nikolic}, \citenamefont
  {Bermudez}, \citenamefont {Martin-Delgado}, \citenamefont {Lewenstein},\ and\
  \citenamefont {Spielman}}]{Goldman2010}%
  \BibitemOpen
  \bibfield  {author} {\bibinfo {author} {\bibfnamefont {N.}~\bibnamefont
  {Goldman}}, \bibinfo {author} {\bibfnamefont {I.}~\bibnamefont {Satija}},
  \bibinfo {author} {\bibfnamefont {P.}~\bibnamefont {Nikolic}}, \bibinfo
  {author} {\bibfnamefont {A.}~\bibnamefont {Bermudez}}, \bibinfo {author}
  {\bibfnamefont {M.~A.}\ \bibnamefont {Martin-Delgado}}, \bibinfo {author}
  {\bibfnamefont {M.}~\bibnamefont {Lewenstein}}, \ and\ \bibinfo {author}
  {\bibfnamefont {I.~B.}\ \bibnamefont {Spielman}},\ }\href {\doibase
  10.1103/PhysRevLett.105.255302} {\bibfield  {journal} {\bibinfo  {journal}
  {Phys. Rev. Lett.}\ }\textbf {\bibinfo {volume} {105}},\ \bibinfo {pages}
  {255302} (\bibinfo {year} {2010})}\BibitemShut {NoStop}%
\end{thebibliography}
\end{document}